\documentclass[aps,twocolumn,showpacs,preprintnumbers,amsmath,amssymb]{revtex4}
\usepackage{graphicx}
\usepackage{dcolumn}
\usepackage{bm}

\newcommand{\bef}{\begin{figure}}
\newcommand{\eef}{\end{figure}}

\newcommand{\be}{\begin{equation}}
\newcommand{\ee}{\end{equation}}
\newcommand{\bea}{\begin{eqnarray*}}
\newcommand{\eea}{\end{eqnarray*}}
\newcommand{\ra}{\rightarrow}

\begin{document}

\title{Thermal photons from $\pi\rho\,\ra\,\pi\gamma$ revisited}

\author{Jan-e Alam$^1$, Pradip Roy$^2$ and Sourav Sarkar$^1$}

\medskip

\affiliation{$^1$Variable Energy Cyclotron Centre, Kolkata 700064, India\\ 
$^2$Saha Institute of Nuclear Physics, Kolkata 700064, India}

\date{\today}
\begin{abstract}
We evaluate the photon spectra from the reaction $\pi\rho\ra\,\pi\gamma$
for the exchange of $\pi$, $\rho$, $\omega$, $\phi$ and $a_1$ as intermediary
mesons. It is found that the contributions from the 
intermediary $a_1$ is more than any other meson exchange processes up to
photon energies 2.5 GeV.

\end{abstract}
\maketitle

Investigation of the properties of hot and dense hadronic matter produced 
in high energy heavy ion collisions through real and virtual photon
spectra is a field of great contemporary interest. 
As the energy density of the hadronic matter increases, the system
is expected to go to a new phase of matter called quark gluon plasma (QGP).
Among others, photons are considered as a very promising
signal of QGP~\cite{annals}. However, to estimate the photons from
QGP an accurate evaluation of the photon spectra from hadrons is
necessary~\cite{phots}. Among all the processes which produces photons in
a hot hadronic system the reaction $\pi\rho\,\ra\,\pi\gamma$ is the
dominant one for photon energies ($E$) above 0.5 GeV.

In a recent paper~\cite{turbide} it was 
claimed that the $t$ channel $\omega$ exchange in 
$\pi\rho\,\ra\,\pi\gamma$ is the single most dominant process
of photon production for $E>2$ GeV.
For $\pi-\rho-a_1$ vertices they have 
employed the Massive Yang-Mills approach and the $\pi-\rho-\omega$
interaction is  similar to that given in~\cite{gsw}. 
It is the purpose of this paper to comment on the above observation.

We have made a detailed study of this process considering all
possible diagrams involving $\pi$, $\rho$, $\omega$, $\phi$ and
$a_1$ mesons in the intermediate state. For this purpose the
following interactions have been considered.
For the $\pi-a_1-\gamma$ and $\pi-a_1-\rho$ vertices we employed 
the phenomenological 
interactions from Ref.~\cite{rudaz}, (see also~\cite{shin}) 
which reproduces the $a_1\ra\,\pi\,\rho$
and $a_1\ra\,\pi\,\gamma$ decay widths reasonably well. It may be mentioned
here that the $a_1\pi\gamma$ vertex 
used in Refs.\cite{song,xiong} gives a
larger value of the above decay width compared to the experimental value.
The $\omega-\rho-\pi$ interaction is taken from ~\cite{gsw}. The
coupling constants has been fixed (via vector meson dominance) 
to reproduce the $\omega\ra\,\pi^0\,\gamma$ decay width. For the sake
of completeness we have also considered $\phi$ mediated reactions,
though its contribution is found to be small. The interaction vertex 
for $\phi-\rho-\pi$ is similar to  $\omega-\rho-\pi$ and the coupling
is constrained from the decay $\phi\ra\,\pi^0\,\gamma$. The $\rho-\pi-\pi$
vertex has been fixed from $\rho\ra\,\pi\pi$ decay. 
All the reactions involving intermediary $\pi$, $\rho$, $a_1$,
$\omega$ and $\phi$ as well as four point $\pi-\rho-\pi-\gamma$ interactions
have been considered. We have not introduced form factors at the vertices
because the main focus of the work is to compare the relative
importance of the $\omega$ and $a_1$ exchange reactions. 
Coherent sums have been performed for the same class of relevant diagrams
for all the intermediary mesons.

The emission rate of photons from $\pi\rho\,\ra\,\pi\gamma$ is plotted
in Fig.~\ref{fig1} at a temperature $T=200$ MeV. Our results do not 
agree with that of Ref.~\cite{turbide} {\it i.e.}  $\omega$-exchange 
in the $t-$ channel is not the single most important process for the
entire range of $E$  considered here. In fact
up to $E\sim 2.5$ GeV the emission rate due to $a_1$-exchange 
processes is seen to be the most dominant one.
We have also observed that the $a_1$ exchange in the $s$ channel 
is more dominant than the corresponding $t$ channel exchange as
claimed in~\cite{xiong}. The contribution from 
$\omega$ exchange processes in the $s$ and $t$ channels 
is found to be
comparable to the $a_1$ exchange processes beyond $E\sim 2.5$ GeV.
The decay of $\omega$ meson is not considered here to avoid
double counting with the $s$ channel $\omega$ exchange process.
Contribution from $\phi$ exchange is negligibly small.

In summary, we have evaluated the production rate of photons by the 
reactions $\pi\rho\ra\,\pi\gamma$ with all the possible charge
states of $\pi$ and $\rho$. An exhaustive set of processes 
involving intermediary  
$\pi$, $\rho$, $\omega$, $\phi$ and $a_1$ mesons have been considered.
We observe that the contributions from the 
$a_1$ exchange process is the dominant one for photon
energies up to 2.5 GeV. It should be mentioned here that beyond
$E=2.5$ GeV photons from hard QCD processes may mask the thermal 
contributions. 

\bef
\begin{center}
\includegraphics[scale=0.45]{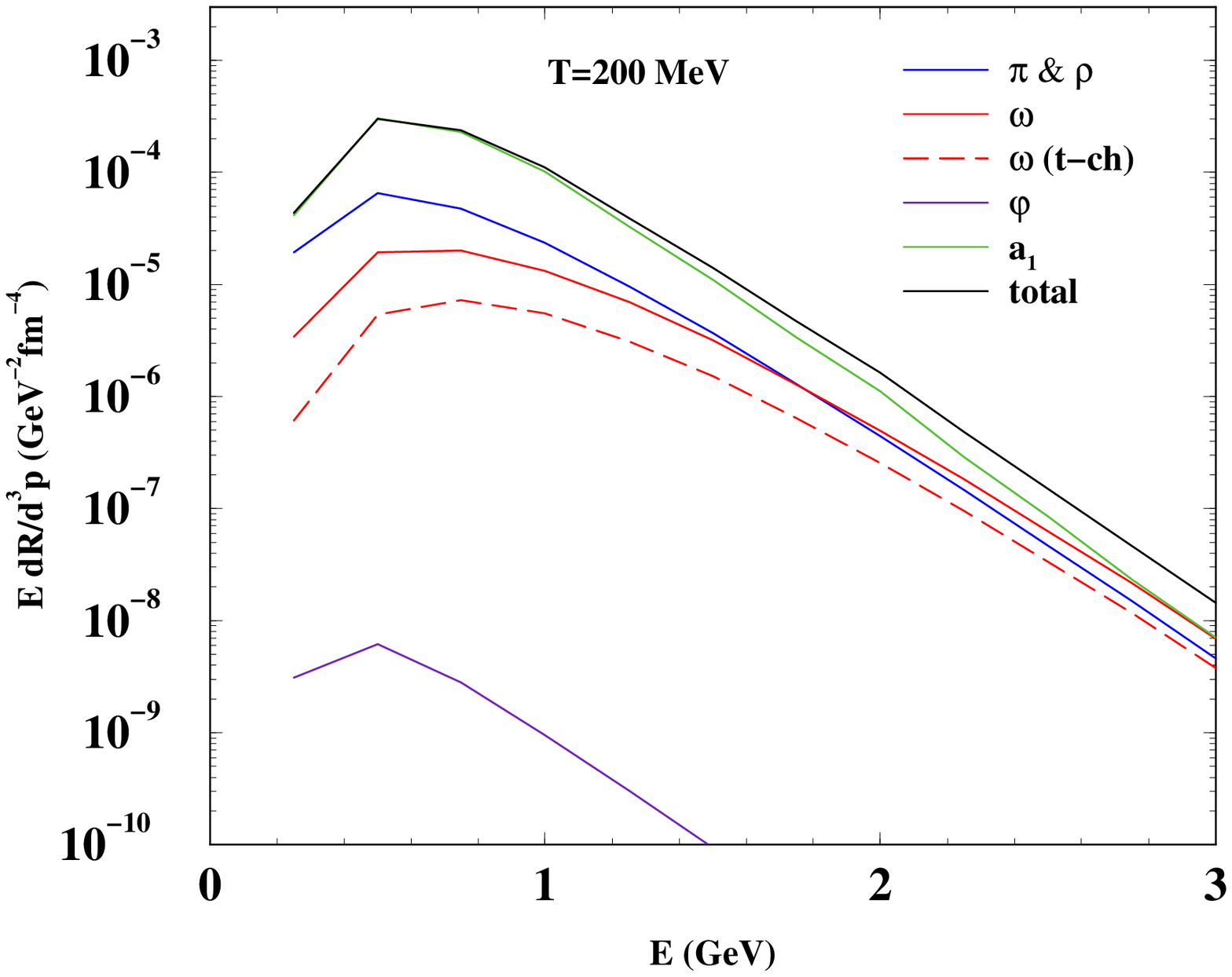}
\caption{Thermal photon emission rates from the reaction
$\pi\rho\,\ra\,\pi\gamma$ at $T$=200 MeV. Red dashed (solid) line
indicates contribution from $\omega$ exchange in $t$ ($s$ +$t$)
channel. Blue (green) line shows the spectra for $\pi$ and $\rho$ 
($a_1$) exchange processes. Contributions from $\phi$ exchange
processes is shown by violet line. The total contribution is
indicated by the solid (black) line.
}
\label{fig1}
\end{center}
\eef

\end{document}